\documentclass{aastex62}
\DeclareGraphicsExtensions{.ps}
\received{April 24, 2019}
\revised{May 24, 2019}
\accepted{June 7, 2019}
\submitjournal{ApJ}

\shorttitle{M~47 white dwarf}
\shortauthors{Richer et al.}

\begin{document}

\title{A MASSIVE MAGNETIC HELIUM ATMOSPHERE WHITE DWARF BINARY IN A YOUNG STAR CLUSTER}

\correspondingauthor{Harvey Richer}
\email{richer@astro.ubc.ca}

\author[0000-0001-9002-8178]{Harvey B. Richer}
\affil{University of British Columbia \\
Vancouver, BC, Canada\\
}

\author{Ronan Kerr}
\affiliation{University of British Columbia \\
Vancouver, BC, Canada\\
}

\author{Jeremy Heyl}
\affiliation{University of British Columbia \\
Vancouver, BC, Canada\\
}

\author{Ilaria Caiazzo}
\affiliation{University of British Columbia \\
Vancouver, BC, Canada\\
}
\nocollaboration

\author{Jeffrey Cummings}
\affiliation{Johns Hopkins University \\
Baltimore, Maryland, USA}
\nocollaboration

\author{Pierre Bergeron}
\affiliation{Universit\'e de Montr\'eal \\
Montr\'eal, Qu\'ebec, Canada}

\author{Patrick Dufour}
\affiliation{Universit\'e de Montr\'eal \\
Montr\'eal, Qu\'ebec, Canada}



\begin{abstract}

We have searched the Gaia DR2 catalogue for previously unknown hot white dwarfs in the direction of young open star clusters. The aim of this experiment was to try and extend the initial-final mass relation (IFMR) to somewhat higher masses, potentially providing a tension with the Chandrasekhar limit currently thought to be around 1.38 M$_{\odot}$. We discovered a particularly interesting white dwarf  in the direction of the young $\sim$150~Myr old cluster Messier~47 (NGC 2422). All Gaia indicators (proper motion, parallax, location in the Gaia colour-magnitude diagram) suggest that it is a cluster member. Its spectrum, obtained from Gemini South, yields a number of anomalies: it is a DB (helium-rich atmosphere) white dwarf, it has a large magnetic field (2.5 MG), is of high mass ($\sim$1.06 M$_\odot$) and its colours are very peculiar --- particularly the redder ones ($r$, $i$, $z$ and $y$), which suggest that it has a late-type companion. This is the only magnetized, detached binary white dwarf with a non-degenerate companion of any spectral type known in or out of a star cluster. If the white dwarf is a cluster member, as all indicators suggest, its progenitor had a mass just over 6 M$_\odot$. It may, however, be telling an even more interesting story than the one related to the IFMR, one about the origin of stellar magnetic fields, Type I supernovae and gravitational waves from low mass stellar systems.

\end{abstract}



\section{Introduction} \label{sec:intro}

Ultramassive white dwarfs have been a topic of considerable research recently (e.g. \citet{2010A&A...524A..36K,2018arXiv180703894C}), as their origins and properties have significant implications with respect to the late stages of stellar evolution. They set the progenitor upper-mass limit to white dwarf production, thought to be in the range of 6-10~M$_\odot$, as well as setting the lower limit to supernova production, which controls much of the energetics and chemical evolution of galaxies (e.g. \citealt{2018ApJ...866...21C}). Knowing the lower-mass limit to the SN II rate also determines the rate of neutron star formation, which helps in establishing the galactic star formation rate, and provides an important test of the physics of dense matter through the Chandrasekhar limit. While some ultramassive white dwarfs have mass estimates of over 1.3~M$_\odot$ (e.g. \citealt{1998AAS...191.1511B,2000ApJ...538..854B}), which is near the Chandrasekhar limit of 1.38~M$_\odot$ \citep{Chandrasekhar}, most of these especially massive objects are better explained through mergers or accretion from a companion because they are more common than the abundance of intermediate-mass stars would indicate \citep{2005ApJS..156...47L}. However, \citet{2016MNRAS.461.2100T} did not find evidence for massive white dwarfs formed from mergers in the solar neighborhood. Regardless, no massive white dwarfs can be clearly attributed to the evolution of a single star except for GD~50 and WD~33 in NGC~2099 \citep{2006MNRAS.373L..45D,2016ApJ...820L..18C,2018ApJ...861L..13G}, both of which are associated with young stellar clusters. As such, the maximum known single-progenitor white dwarf mass is still well below the Chandrasekhar limit (see \citealt{2016ApJ...820L..18C}). 

\begin{figure}[h]
    \centering
    \includegraphics[width=11cm,clip,trim=0 2in 0 1in]{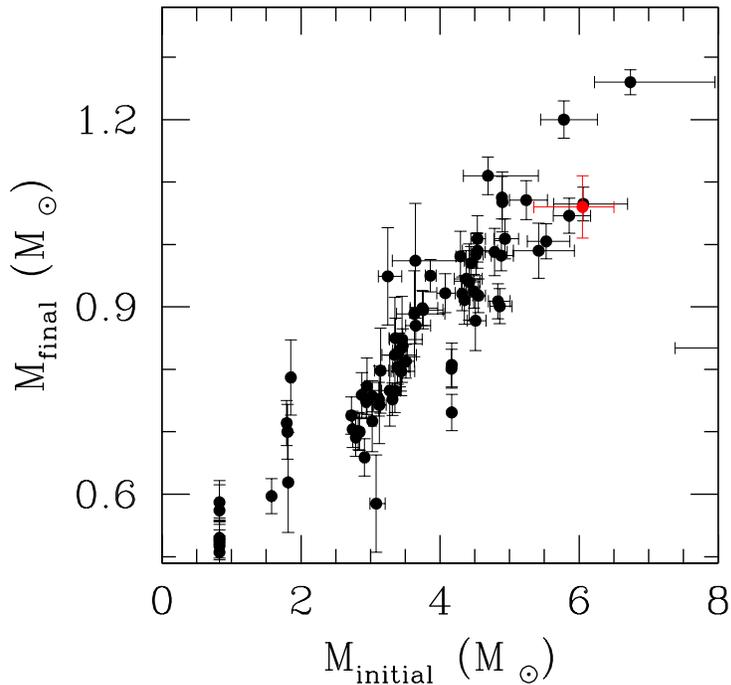}
    \caption{The present status of the observational IFMR \citep{2018ApJ...866...21C}. Note that the current data do not as yet yield white dwarfs near the Chandrasekhar limit that forms the upper edge of the figure. The origin of the red point will be discussed in the following sections.} 

    \label{fig:ifmr}
\end{figure}

GD~50, a member of the AB~Doradus moving group and possible former Pleiades member \citep{2018ApJ...861L..13G}, is the most massive known single-progenitor white dwarf at 1.26$\pm{0.02}$ M$_{\odot}$, having an age and origin consistent with a progenitor mass of 6.74$^{+1.21}_{-0.52}$~M$_{\odot}$ \citep{2018ApJ...866...21C}. Before the discovery of GD~50, it was unclear whether single progenitors could create white dwarfs more massive than $\sim$1.1~M$_{\odot}$, and therefore its discovery greatly increased the range of systems in which we could expect to find white dwarfs. Figure~\ref{fig:ifmr} displays the current state-of-the-art for the observational IFMR, with most of the data taken from a recent compilation \citep{2018ApJ...866...21C}. \citet{2012ApJ...746..144Z} and \citet{2018ApJ...860L..17E} demonstrate two alternative empirical techiques using wide binaries and field stars respectively. More theoretically based versions of the IFMR can be found in \citep{2007A&A...469..239M,2015MNRAS.446.2599D,10.1093/mnras/sty1925} that explore the metallicity dependence as well.

Young open clusters provide an ideal environment in which to search for single-progenitor white dwarfs. In these environments, cluster ages can be derived by fitting isochrones to the stellar population, providing a time of formation for any associated objects. As such, a white dwarf with a derived cooling age can be compared to the age of the cluster itself, providing a possible lifespan for the progenitor and hence its mass. If such an explanation produces a reasonable result, then merger-based explanations are unnecessary. By surveying clusters of the age of the Pleiades and younger, any white dwarfs found should have progenitor masses similar to or greater than that of GD~50.

\section{GAIA SEARCH} \label{sec:search}

Using Gaia DR2 data \citep{2018A&A...616A...1G}, we searched 
broad fields around all open clusters in the WEBDA catalog \citep{1995ASSL..203..127M} with turnoff masses between 
4.5~M$_\odot$ and 15~M$_\odot$ (ages between 158 and 10~Myrs), and distances less than 800~pc. There are a total of 95 catalogued open clusters that satisfy these criteria. We located a previously unknown white dwarf in the direction of 
M 47 (NGC 2422), a cluster at a distance of 501~pc from the Sun, of total mass about 450~M$_\odot$ and with a young age of 150$\pm 20$ Myrs \citep{2018AJ....156..165C, 2005A&A...438.1163K}. The Gaia catalogue entry for this star is 3029912407273360512 and its coordinates are 
RA:114.0287$^{\circ}$, Dec:$-$14.8586$^{\circ}$ (epoch 2015.5).  A cluster as young as M~47 is expected to have a current turnoff mass around 5.4~M$_\odot$. Hence if the white dwarf is unambiguously a member of the cluster, we expect that the mass of the star that produced it would be comparable to or perhaps even greater than that of the precursor to GD~50. It is important to note that the object discovered here is not the same one discussed in \citet{1981A&A....99L...8K}. While that star is also in the general direction of M~47, its Gaia distance is half that of the cluster and its proper motion does not agree with the bulk cluster motion. 

\begin{figure}
    \centering
    \includegraphics[width=0.45\textwidth,clip,trim=0in 2.5in 0.2in 0.4in]{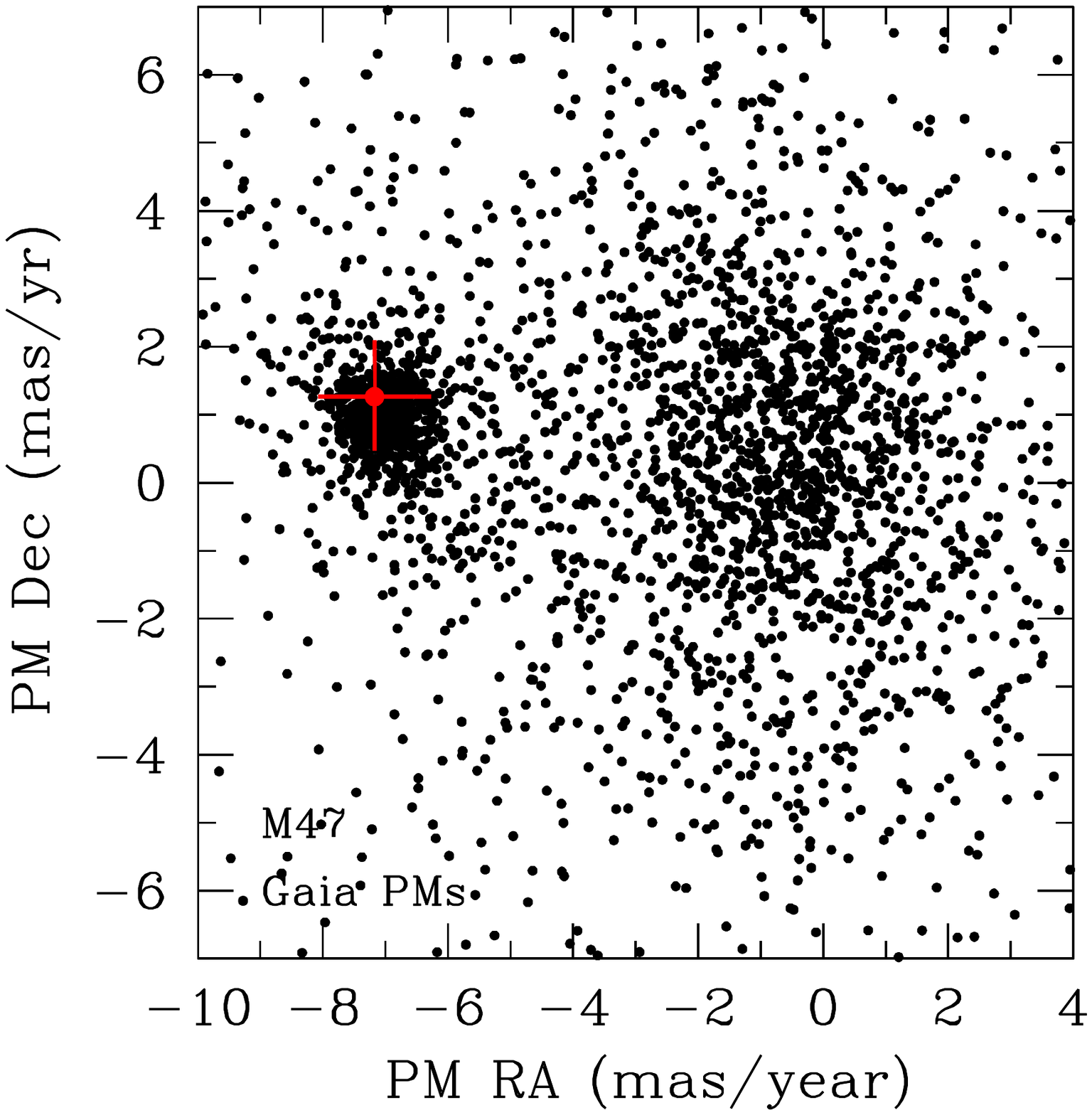}
       \includegraphics[width=0.45\textwidth,clip,trim=0in 2.5in 0.2in 0.4in]{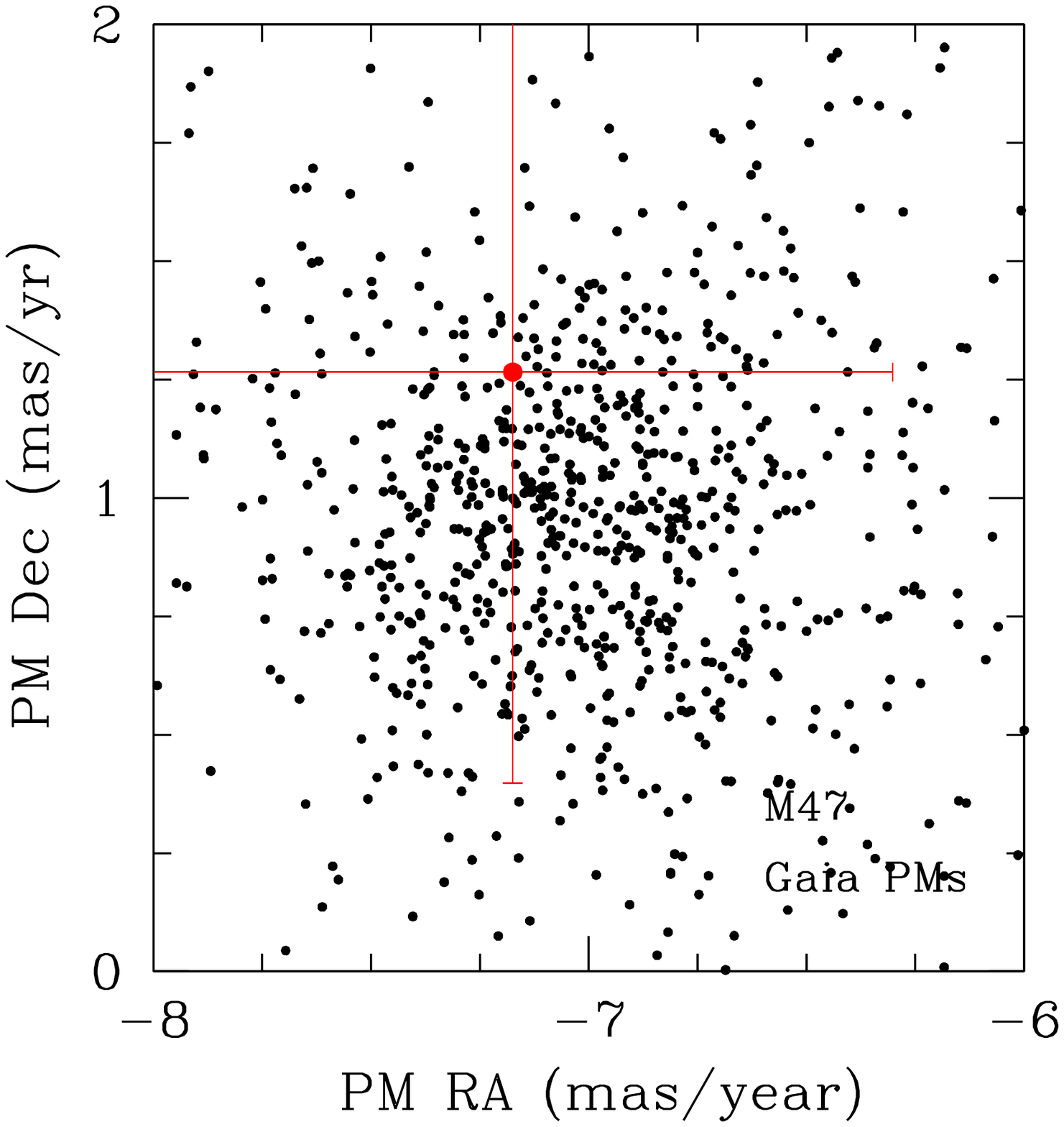}
    \caption{Left: The Gaia proper motions in a 2 by 2 degree field that contains M~47. The galactic field is the large blob centered near (0,0) while M~47 is the tighter grouping to the left. The potential M 47 white dwarf is plotted with a red dot.  Right: The proper motions in an expanded plot centered on M~47 with the white dwarf again plotted as a red dot. In both plots the $\pm 1\sigma$ error bars are indicated.}
    \label{fig:pms1}
\end{figure}

\begin{figure}
    \centering
    \includegraphics[width=0.418\textwidth,clip,trim=0in 1.75in 0.2in 0.4in]{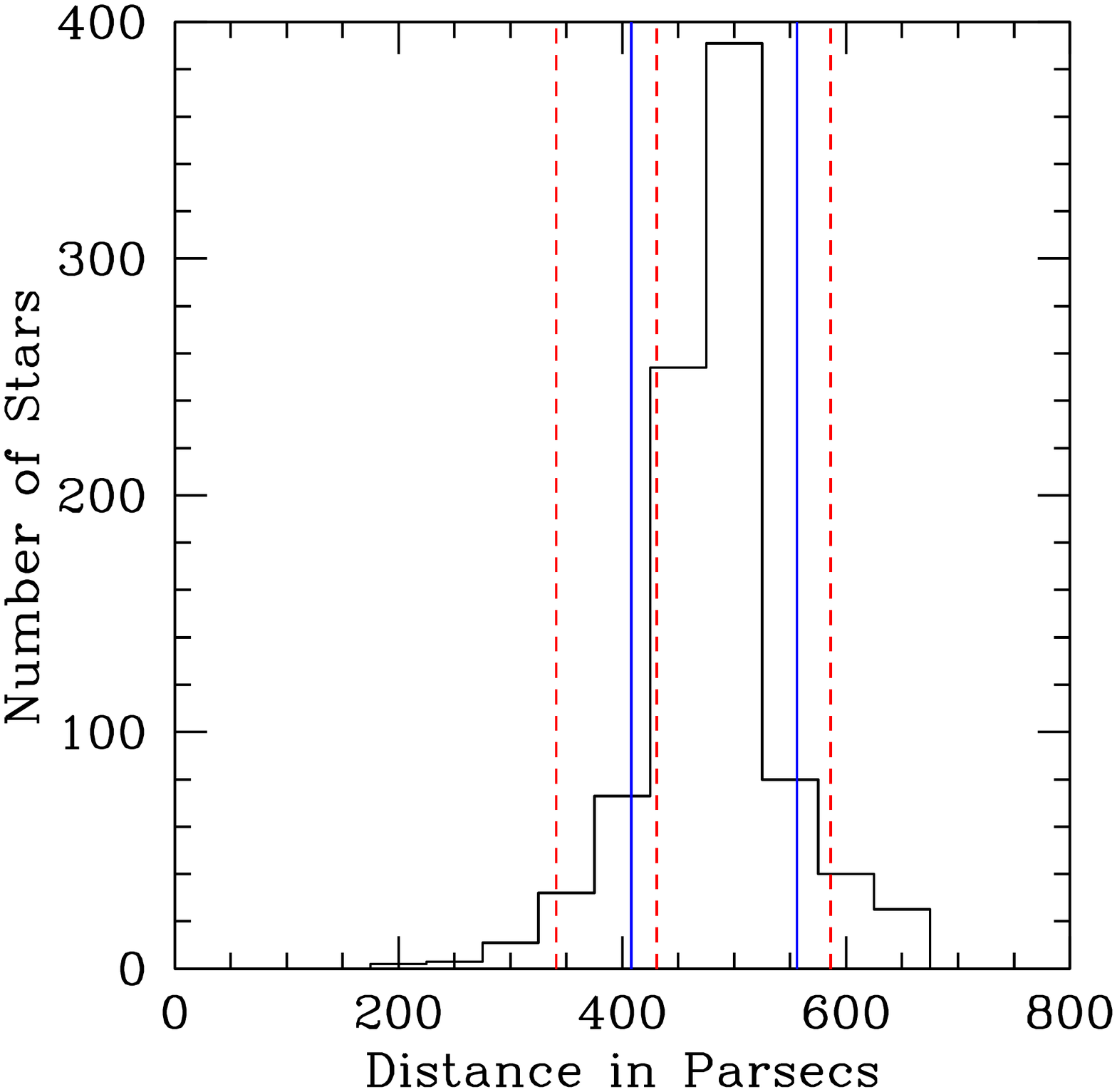}
       \includegraphics[width=0.48\textwidth,clip,trim=0in 2.5in 0.2in 0.4in]{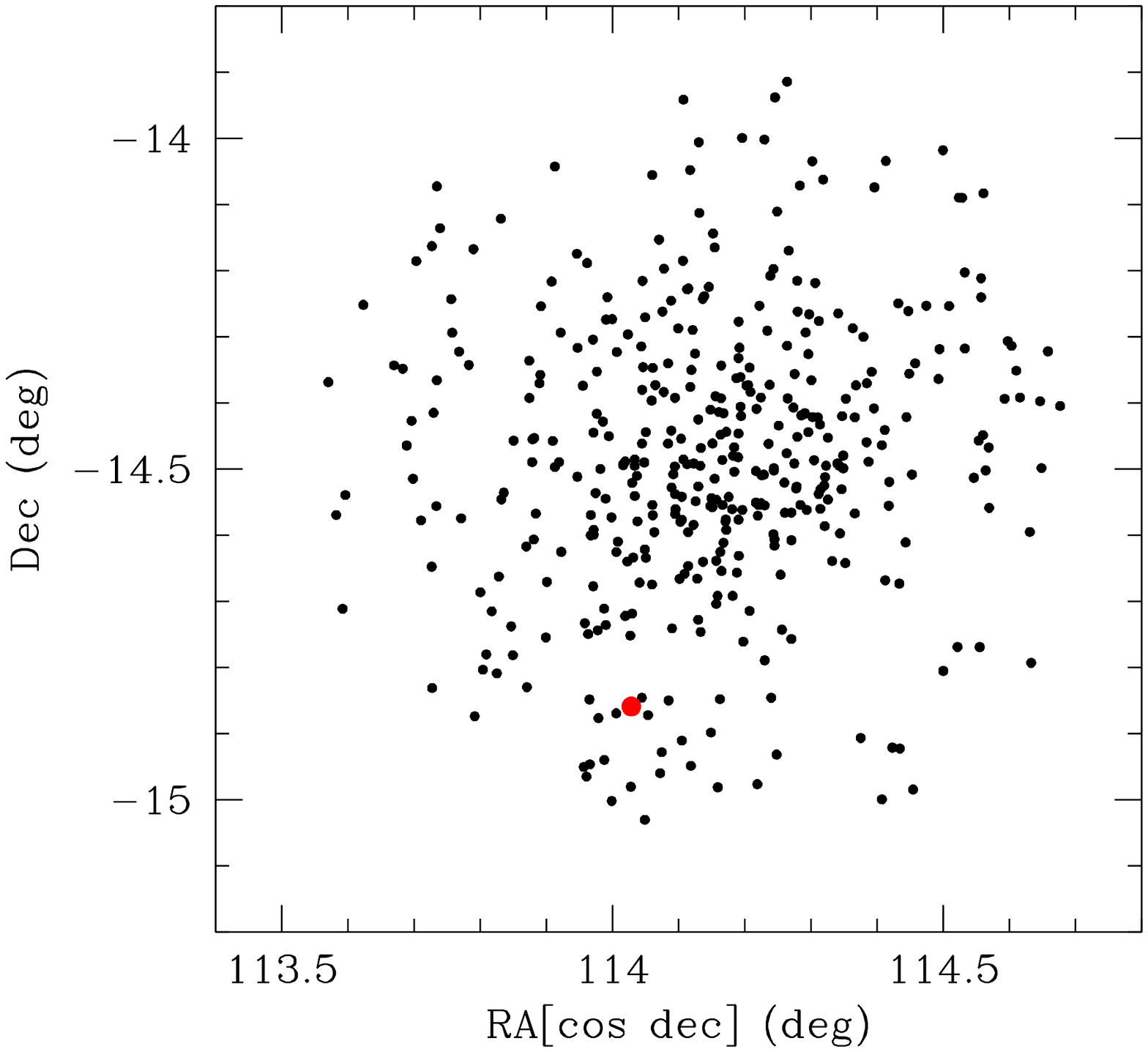}
    \caption{Left: The distribution of parallaxes in the direction of M~47 after a proper motion cut. The middle red dashed line indicates the parallax-derived distance of the white dwarf, and the left and right red dashed lines indicate the one-sigma error range for the distance of the white dwarf. The blue solid lines indicate the $\pm 1\sigma$ parallax error range for stars that pass the proper motion cut. Right: The Gaia proper motion and parallax-selected cluster members centered on the location of
    M~47 on the sky. The M~47 white dwarf is plotted with a red dot.}
    \label{fig:pms2}
\end{figure}

Figure~\ref{fig:pms1} shows the Gaia proper motions in the direction of M~47 with the right panel being a zoom into the cluster center.  The putative white dwarf is plotted with a red dot in both panels. In proper motion space, the white dwarf is well within the cluster distribution:
its proper motion is less than 1$\sigma$ away from that of the cluster center but its error is large as it is among the faintest objects in these plots. In Figure~\ref{fig:pms2}, the left panel illustrates the distribution of parallaxes in the direction of the cluster. The blue lines are the 1$\sigma$ error range in the parallax while the red lines indicate the parallax of the white dwarf (central line) together with its 1$\sigma$ errors. The right panel displays the proper motion and parallax selected M~47 cluster members projected onto the sky. 

\begin{figure}
    \centering
     \includegraphics[width=0.45\textwidth]{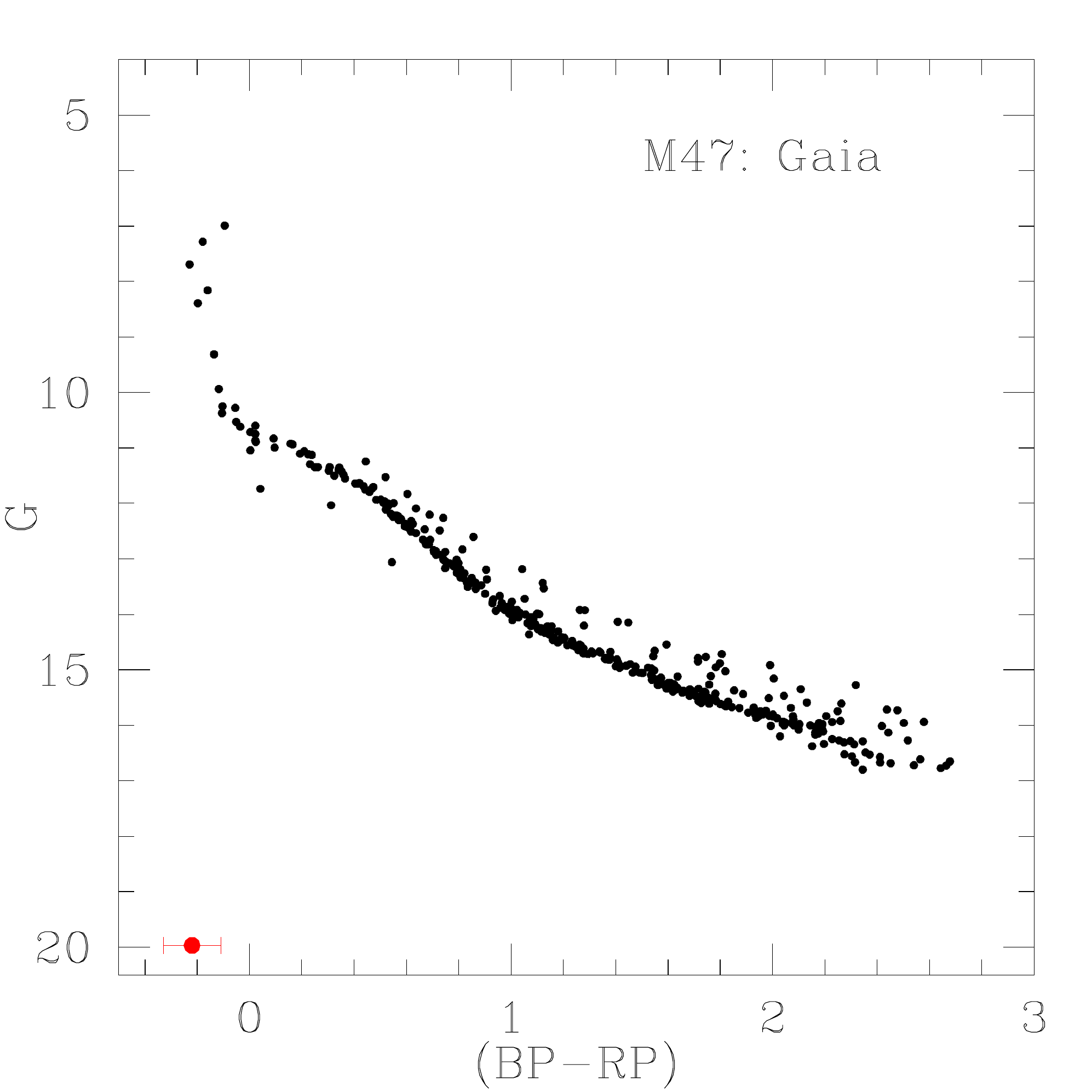}
      \includegraphics[width=0.45\textwidth]{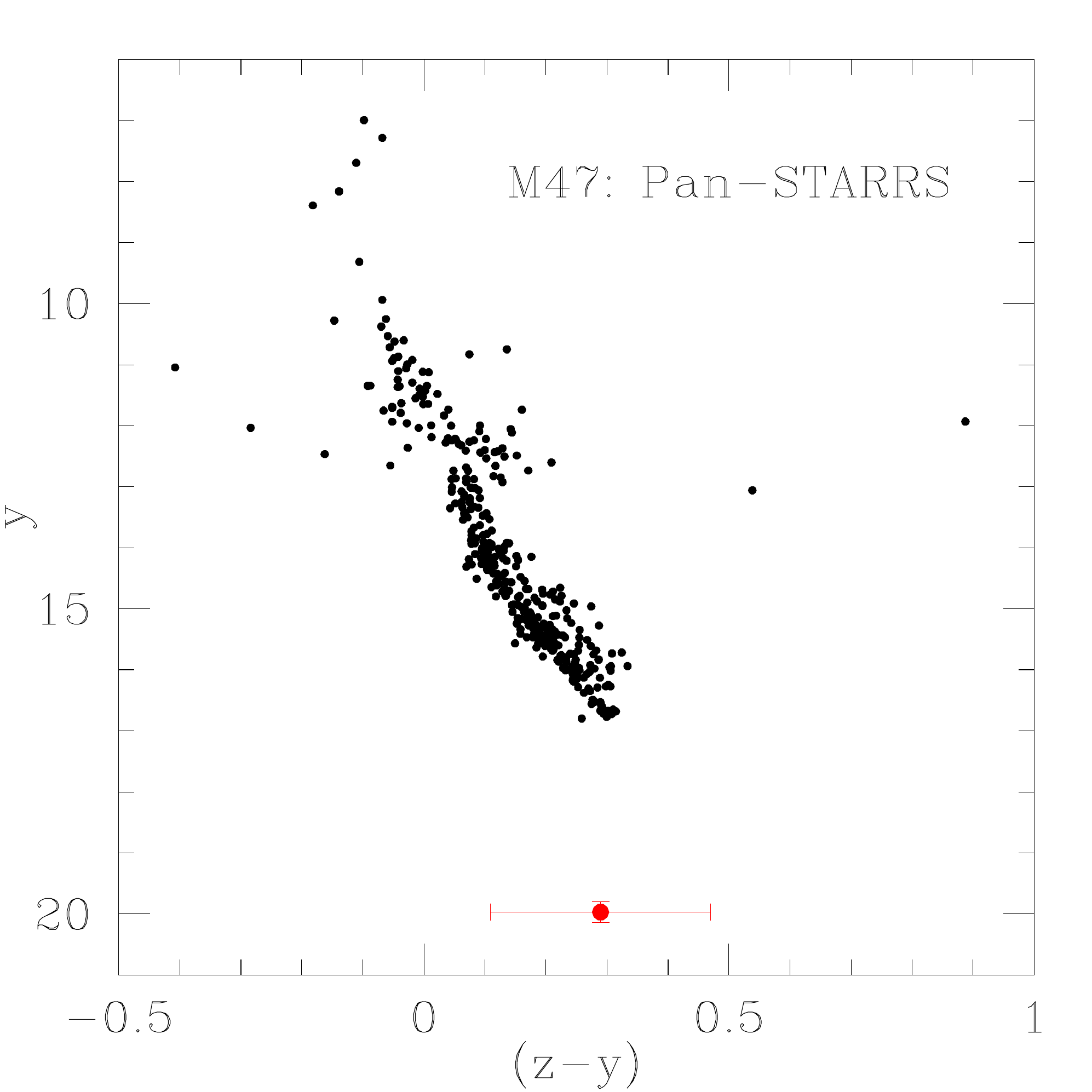}
    \caption{Left: The Gaia CMD of M~47 after proper motion, parallax and color excess cuts designed to isolate the cluster stars. The cluster white dwarf is seen as the red point at the lower left of the diagram. Right: The M~47 CMD in the PanSTARRS filters y and z with the same selections on the Gaia data. The M47 white dwarf is again plotted as a red point and it is interesting that it now almost appears as an extension of the main sequence as its M dwarf companion dominates the light at these longer wavelengths.}
    \label{fig:cmd}
\end{figure}

Figure~\ref{fig:cmd} (left) shows our Gaia-derived color-magnitude diagram of cluster members in M~47 identified with cuts in parallax (1.900 to 2.384 mas), proper motion (1 mas/year radius centered on (-7.0, 1.0)) and a Gaia color excess cut $<$1.5. The candidate white dwarf is seen in the lower left corner with a $\pm 1\sigma$ error bar. The Gaia photometry of 20th magnitude objects is uncertain, and therefore this figure should be used cautiously when attempting to draw precise conclusions about the nature of the object. In figure~\ref{fig:cmd} (right) we plot the PanSTARRS \citep{2016arXiv161205560C} color-magnitude diagram of the cluster in y, (z - y) for the same stars plotted in the Gaia-based diagram. The white dwarf now occupies a rather peculiar location, almost on an extension of the main sequence. As we will see shortly, the white dwarf has a red companion so this is not surprising.

The photometry (with associated errors) of the white dwarf in the Pan-STARRS DR1 as well as the VPHAS+ catalogues \citep{2014MNRAS.440.2036D} is contained in Table 1, and, as can been seen here, the observed magnitudes in the various surveys are quite consistent, allowing us to pursue the white dwarf's properties with some confidence.

\begin{deluxetable}{l|ccc}[h]
\tablecolumns{4}
\tabletypesize{\small} 
\tablecaption{Photometry of M~47 WD}
\tablewidth{0pt}
\tablehead{
 & \colhead{GaiaDR2}&\colhead{Pan-STARRS}\  & \colhead{VPHAS+}
}
\startdata
G   & 19.796 $\pm $ 0.005 & .... & .... \\
(G$_{\rm{BP}}$-G$_{\rm{RP}}$) & -0.13 $\pm $ 0.11 & .... & .... \\
u & .... & .... & 19.58 $\pm $ 0.03 \\
g & .... & 19.75 $\pm $ 0.01 & 19.81 $\pm $ 0.02 \\
r & .... & 19.91 $\pm $ 0.03 & 19.87 $\pm $ 0.03 \\
i & .... & 20.13 $\pm $ 0.02 & 20.05 $\pm $ 0.05 \\
z & .... & 20.25 $\pm $ 0.05 & .... \\
y & .... & 19.97 $\pm $ 0.17 & ....\\
\enddata
\space
\tablenotetext{a}{All these magnitudes are apparent AB magnitudes and are listed before any reddening or extinction corrections are applied.}
 
\end{deluxetable}

The distance and extinction in the direction of the cluster will be critical in determining the white dwarf's properties.
Our Gaia-derived cluster distance is 472$^{+85}_{-63}$ pc (Figure \ref{fig:pms2}), which is in excellent agreement with the value of 501$^{+29}_{-27}$ pc derived by \citet{2018AJ....156..165C} using model fits to the observed UBV CMD. They have also shown that the cluster's reddening is $E(B-V) = 0.065$. Using data from Leo Girardi's web page (http://stev.oapd.inaf.it/cgi-bin/cmd), we derive $A_G/A_V = 0.862$ hence $A_G = 0.174$ and thus $M_G = 11.122$. Also, $E(BP - RP) = (1.07198A_V - 0.64648A_V)$ yielding an intrinsic colour for the white dwarf of $(G_{BP} - G_{RP}) = -0.219$. Using the tables of extinction values in \citet{2013MNRAS.430.2188Y} we derive the
absolute magnitudes and intrinsic colors of the M~47 WD in the VPHAS+ and PanSTARRS systems collected in Table 2.

\begin{deluxetable}{l|ccc}[h]
\tablecolumns{4}
\tabletypesize{\small} 
\tablecaption{Absolute Magnitudes and Intrinsic Colours of M~47 WD}
\tablewidth{0pt}
\tablehead{
 & \colhead{GaiaDR2}&\colhead{Pan-STARRS}\  & \colhead{VPHAS+}
}\startdata
M$_G$   & 11.22 & .... & .... \\
(G$_{\rm{BP}}$-G$_{\rm{RP}}$)$_0$\ & -0.22 & .... & .... \\
M$_{\rm{u}}$ & .... & .... & 10.79 \\
M$_{\rm{g}}$ & .... & 11.04 & 11.10 \\
M$_{\rm{r}}$ & .... & 11.26 & 11.22 \\
M$_{\rm{i}}$ & .... & 11.52 & 11.44 \\
M$_{\rm{z}}$ & .... & 11.67 & .... \\
M$_{\rm{y}}$ & .... & 11.45 & ....\\
\enddata
\end{deluxetable}


\section{The Spectrum of the white dwarf} \label{sec:wdspectrum}

\begin{figure}
    \centering
    \includegraphics[clip,trim=0.3in 0 0 0]{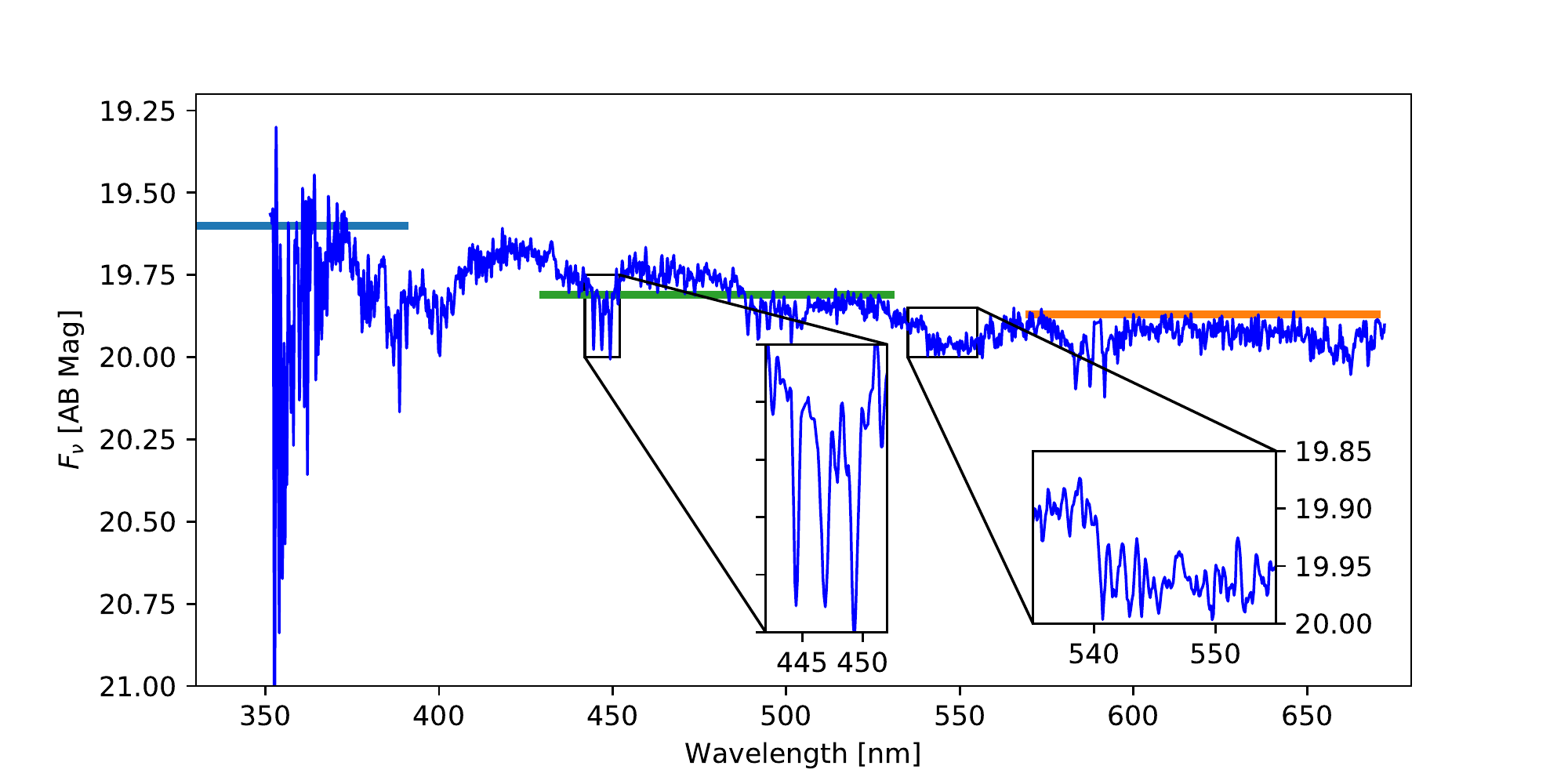}
    \caption{Spectrum of the M~47 white dwarf. It exhibits only He lines, all of which are Zeeman split by about 0.014~eV (see particularly the 447.1~nm line) due to the object's high magnetic field. The insert at 540~nm displays what is thought to be a TiO band contributed by a cool companion. The photometric bands u, g and r are indicated with horizontal lines covering the wavelength range of each band and placed at the appropriate vertical level of its photometry in PanSTARRS. Since the spectrum was fluxed independently of the photometry, it is clear that they are in good agreement with each other.}
    \label{fig:spectrum}
\end{figure}

In order to try and better constrain the nature of the object as a white dwarf and to derive its properties more precisely, we obtained a spectrum of the white dwarf with GMOS in long-slit mode on Gemini South, exploiting its fast turnaround capability. We used the B600 grating centered at 512~nm with no blocking filter. The slit width, set to $1"$, provided about 1{\AA} resolution. The total exposure time on the object was 1.8 hours. The results of the spectroscopic observations turned out to be quite a surprise. 
Figure~\ref{fig:spectrum} illustrates the unusual spectrum of the white dwarf in M~47. 

From the spectrum, we were able to conclude the following:
\begin{enumerate}
    \item The object is a DB white dwarf -- that is, a star with a pure helium atmosphere. Note that the spectrum only contains He lines (e.g.\ 447.1~nm) and no hydrogen Balmer lines.
    \item All the lines are triplets, indicating Zeeman splitting by a high magnetic field. \emph{This is the only DB cluster white dwarf known to have a magnetic field}. From modelling the amount of Zeeman splitting (0.014~eV), we estimate the field strength to be $\sim 2.50\pm0.38$~MG. The field and its associated error were derived from cross-correlating the spectrum with spectra shifted in energy to measure the splitting.
    \item The $(u-g)_0$ colour of the star ($-0.31$ in AB magnitudes or $-1.30$ in VEGAMAGS) indicates that the star is very hot with an effective tempersture in excess of 25,000~K. 
    \item There is a peculiar feature seen in the spectrum that has the appearance of a molecular band head. Located at $\sim$540~nm, its position agrees well with a TiO band commonly seen in M dwarf stars.
\end{enumerate}
\vspace{0.1cm}

\section{Derived Properties of the white dwarf} \label{sec:wdproperties}

\begin{figure}
    \centering
    \includegraphics[clip,trim=0.3in 0 0 0 ]{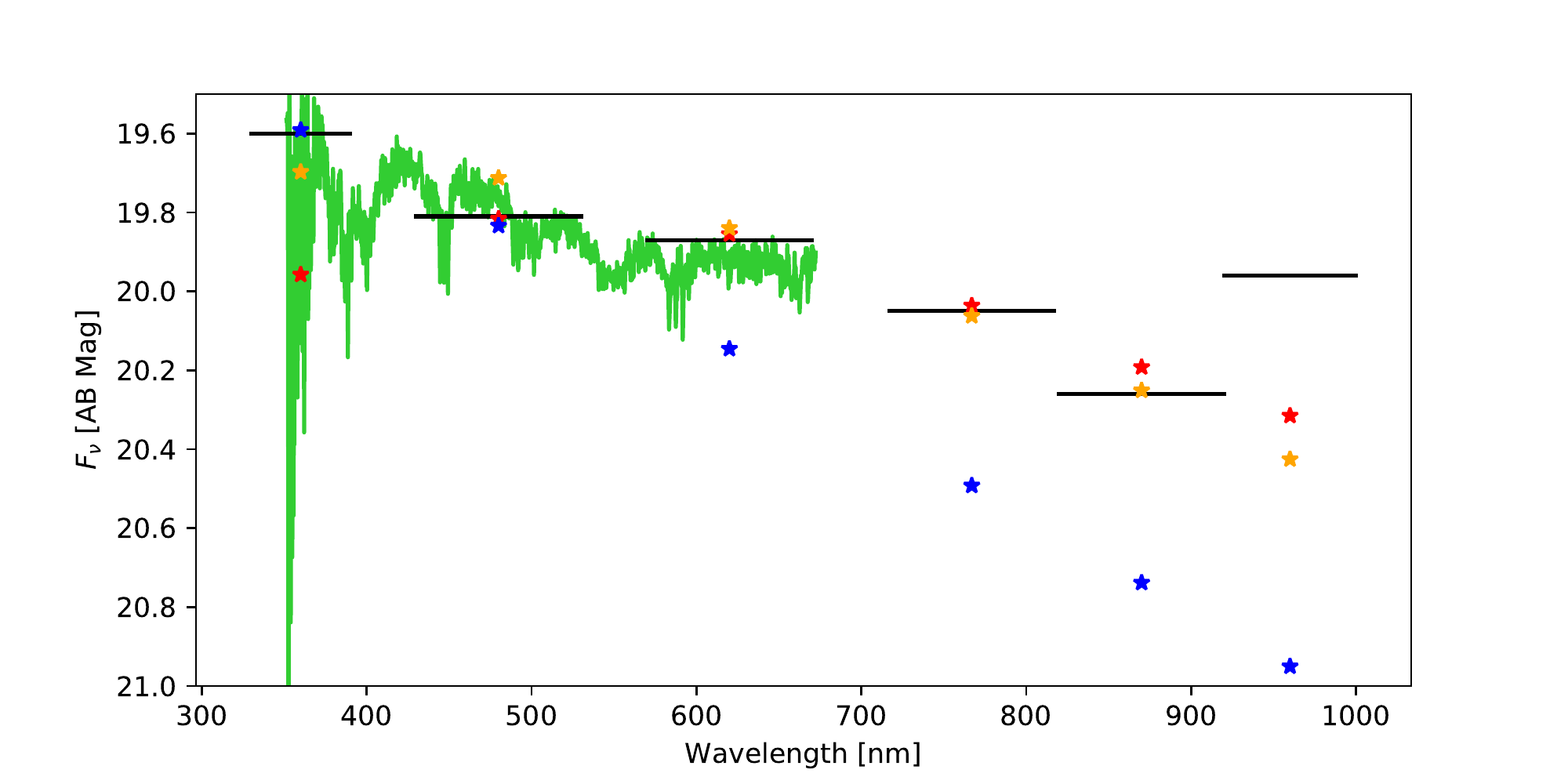}
        \caption{Spectrum of the M~47 white dwarf together with all the available photometry and some theoretical model data. The horizontal lines correspond (from the left) to the bandpasses and magnitudes of $u$, $g$, $r$ and $i$ from VPHAS+ and $z$ and $y$ from PanSTARRS. The blue, yellow and red stars are model magnitudes (with extinction and distance modulus applied) for DB white dwarfs at temperatures of 31,000~K (which fits the $u$ and $g$ magnitudes), 13,000~K (which fits the $r$, $i$ and $z$ magnitudes), and 11,000~K (which is a better, but not very good fit at $z$ and $y$.)}
    \label{fig:spectrum1}
\end{figure}

We attempted to fit the stellar photometry to pure DB white dwarf models of various masses\footnote{\url{http://www.astro.umontreal.ca/~bergeron/CoolingModels}} as shown in Figure~\ref{fig:spectrum1}.
All the models were placed at a distance of 501~pc with the interstellar reddening in the direction of the star, $E(B-V) = 0.065$, taken into account. It was immediately clear that it was not possible to fit the entire spectrum with a single white dwarf model.

Using the available VPHAS+ $u$ and $g$ magnitudes together with the associated errors, we applied this photometry to the Montreal white dwarf cooling models for pure helium atmosphere white dwarfs$^1$.  This determines its T$_{\rm eff}$, mass, and cooling age, and the propagation of these photometric errors translates to the errors in these three parameters.  Application of this cooling age and its errors to the photometrically fit age of 150$\pm$20 Myr for M 47 (Cummings \& Kalirai 2018) using the MIST isochrones (Choi et~al.\ 2016) gives the progenitor's evolutionary timescale and its errors.  Lastly, the application of this timescale back into the MIST isochrones provides the initial progenitor mass, and its errors, that takes this timescale to reach the white dwarf cooling sequence.

\begin{figure}
    \centering
     \includegraphics[width=0.59\textwidth,clip,trim=0in 2in 0.2in 0.4in]{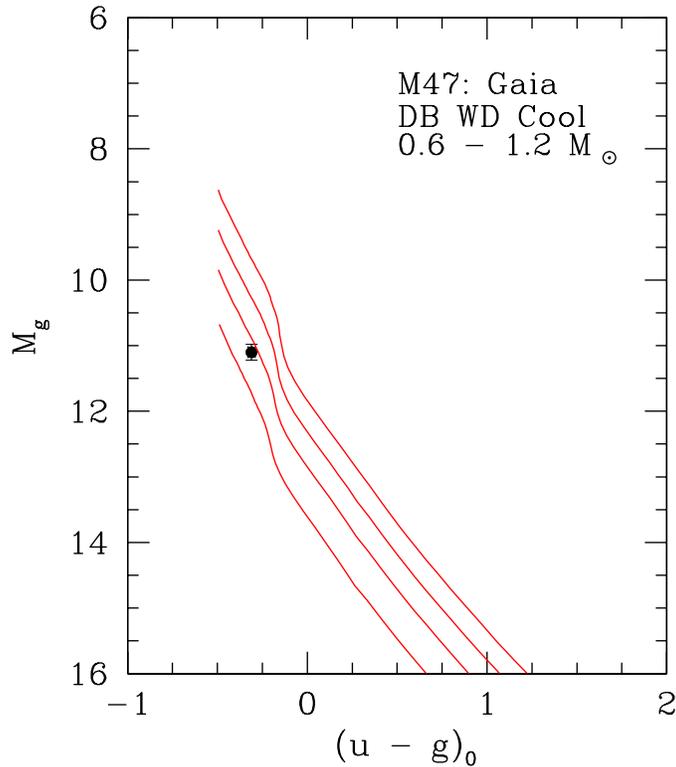}
    \caption{The M$_{g}$, (u-g)$_0$ CMD containing just the white dwarf and DB cooling models for masses ranging from 
    0.6~M$_{\odot}$ at the top to 1.2~M$_{\odot}$ at the bottom. The small error bars include photometric errors as well as the uncertainty in the cluster distance.} 
    \label{fig:cmd1}
\end{figure}

The model best agreeing with these data is a 1.06$\pm 0.05$~M$_{\odot}$ model with a temperature of 32,100$\pm 2700$~K and a white dwarf cooling age of 76$\pm 15$~Myr (see Figure~\ref{fig:cmd1}). This yields an initial mass for the white dwarf of 6.05$^{+0.70}_{-0.45}$~M$_{\odot}$, providing a reasonable match to the cluster properties. However, in Figure~\ref{fig:spectrum1}, clearly the $r$, $i$, $z$ and $y$ photometry do not fit the 32,100~K-1.06~M$_\odot$ white-dwarf model at all. The star is bright in the $u-$ and $g-$bands (demanding a high temperature) but it has significant flux in the redder bands as well, suggesting a lower temperature. Both of these apparently contradictory properties can be accounted for if the system is a binary with a hot DB white dwarf as the primary and an M dwarf as the secondary
\citep{2002AJ....123..430S,2003AJ....125.2621R}. This nicely accounts for all the observed properties of the system, including the molecular band seen in the spectrum at $\sim$540~nm.  There is no evidence of interaction between the two stars in the binary system as no emission lines are seen in the spectrum. Also, the lack of hydrogen absorption lines seems to exclude a recent accretion episode. We cannot exclude the cool companion from being simply in the line of sight, but that is of rather low probability. We examined the VPHAS+ images, searching for any extension in the stellar profile, but nothing obvious was found.  If the system indeed consists of a white dwarf and an M dwarf, in the $i$-band filter the M dwarf will contribute about one-third of the flux; while in $u$-band, it hardly contributes at all.  By examining the VPHAS+ astrometry in the various bands we can constrain the $i$-band position to lie within 0.16~arcseconds of the $u$-band position, and the separation of the two stars to be less than 0.5~arcseconds (or 250~AU at the distance of M~47).  Given the density of stars in the $i-$band down to the magnitude of the potential M dwarf (0.004 stars per square arcsecond), we would expect only 0.003 stars to lie at least this close to the white dwarf by chance, so it is very likely that the two stars are truly associated.


If we assume the system to be a binary, the $g$- and $u$- photometric bands are the least affected by the M dwarf companion, and we can use them to fit a model for the white dwarf. In Figure~\ref{fig:cmd1}, we plot the white dwarf on a CMD in the $g$- and $u$-bands, together with cooling models for non-magnetic DB white dwarfs of 0.6, 0.8, 1.0 and 1.2~M$_\odot$. If we interpolate in mass between the models, we derive a mass for our white dwarf of 1.06$\pm 0.05$~M$_{\odot}$. Plotting this as M$_{final}$ and using 6.05~M$_{\odot}$ as M$_{initial}$, we locate the M~47 white dwarf in Figure~\ref{fig:ifmr}. It is indicated with a red point. It should be noted that the companion red dwarf could have some effect on the system's $g$ magnitude. This would cause us to underestimate photometrically both the final mass and temperature, potentially making this white dwarf more interesting rather than less.
\hfill \break

\section{Is the White Dwarf a Cluster member?} \label{sec:wdmember}

A critical question that we have not addressed in detail thus far is whether the white dwarf is statistically likely to be a member of M~47. The properties derived from the Gaia data strongly suggest cluster membership, while those from the photometry are less conclusive. We can provide some statistics on cluster membership by investigating the totality of white dwarfs in the Gaia data set. 

\citet{2019MNRAS.482.4570G} provide a very useful and the most
complete catalogue available of white dwarf stars found in the Gaia DR2 survey. We use their compilation to provide statistics on the probability of cluster membership. The entire white dwarf population in Gaia DR2 numbers 486,662 stars. Of these, 565 have proper motions and parallaxes within 1$\sigma$ of those of M~47. In the 40,000 square degrees of sky covered by Gaia, this yields 0.0143 stars per square degree with the astrometric properties of M~47, so there is a 1.4\% chance of having a random white dwarf projected within one square degree of the cluster possessing the cluster properties of distance and proper motion.

This suggests that membership is probable but not overwhelmingly so. Because the field around the cluster is so busy, and the chance of a random alignment is at the 1\% level, it certainly cannot be excluded that the M~47 white dwarf is only seen in projection in the direction of the cluster and is not physically associated with it. The key deciding factor here would be if the physical properties of the white dwarf itself were consistent with those of the cluster and general white dwarf properties.   In the hundred-square-degree region surrounding M~47, only seven percent of white dwarfs with measured temperatures are hotter than 25,000~K (fewer than half of the white dwarfs have temperature estimates).   A white dwarf that is consistent with the age of the cluster must be hotter than this. This additional constraint reduces the chance of a random superposition to less than one chance in a thousand.  If we re-examine Figure~\ref{fig:ifmr} again, and now note that the white dwarf in M~47 is included in this diagram plotted as the red point seen in the upper right. Its location conforms reasonably well to the IFMR, even if it is somewhat lower in mass than would be expected for its initial mass. It is not the only one though, as there are quite a few other white dwarfs in the same area of the diagram. At these higher masses, this larger scatter has been shown to be beyond that expected by errors alone, but it may be explained by differences in progenitor rotation rates \citep{2019ApJ...871L..18C}.

The white dwarf's location in the IFMR diagram, its Gaia astrometric properties compared to those of the cluster, its age and initial mass both being consistent with cluster membership, all together make for a very strong case that the DB white dwarf is indeed a cluster member. 

\section{Some Concluding Comments on this Remarkable white dwarf} \label{sec:conclusion}

White dwarfs in young clusters, other than providing constraints on the upper end of the IFMR, provide information about the last evolutionary stages of intermediate-mass stars. Finding a high mass and highly magnetized DB white dwarf as the first white dwarf produced by a young cluster is both surprising and exciting as it has implications on the theory of the origin of magnetic fields in white dwarfs. In particular, high magnetic fields are thought to be either the product of a merger or to be produced by powerful dynamos in the convective cores of intermediate-mass progenitors \citep{2016Natur.529..364S}. Our finding is possibly a confirmation of the dynamo theory, as we know the progenitor's mass to be higher than 6~M$_\odot$. 

It will be important to measure the orbital period of the system. This will tell us whether the stars are likely to be interacting, although the lack of emission in the spectrum already suggests that they are not, or whether they will merge in a Hubble time. The combined mass of the system is potentially up to 1.5~M$_{\odot}$ ($\sim1.1$~M$_{\odot}$ for the white dwarf and $\sim0.4$~M$_{\odot}$ for the M dwarf companion), which does exceed the Chandrasekhar limit. The system is thus a possible progenitor of a type Ia supernova.  

The magnetic field of the white dwarf is lower than 3~MG. This argues that the system is unlikely to be a magnetic cataclysmic variable (MCV) that is currently in a state of low or no accretion \citep{2015SSRv..191..111F}. If the system does, however, turn out to be an MCV, it would be the first system of this type found in such a state.
Alternately, if the orbital period is measured to be long enough that the two stars never come into contact during a Hubble time, then the system would be the first magnetic white dwarf known with a detached non-degenerate companion \citep{2005AJ....129.2376L,2018arXiv181006146F,2018MNRAS.478..899B}.  Among non-magnetic white dwarfs, the occurrence of detached binaries is $\sim$20\%. With the possible exception of this M~47 white dwarf, no magnetic white dwarfs have yet been found in detached binaries with a non-degenerate companion. \citet{2015ApJ...804...93L} used this statistic to strongly suggesting that the merger and accretion event produce magnetic fields in these objects \citep{2015SSRv..191..111F,2018CoSka..48..228K,2018MNRAS.478..899B}. The M~47 white dwarf stands out as a counter-example to this scenario, likely because the precursor and the white dwarf itself are so massive.

The strong magnetic field of the white dwarf, particularly if it increases towards the center of the star, can distort its shape. If it is also rapidly spinning, it could be a source of gravitational radiation \citep{2000MNRAS.317..310H, 2019arXiv190502730K}

What is even more remarkable in this finding is that the magnetic primary is also a DB white dwarf and that the system lives in a young star cluster. \citet{2019MNRAS.482.5222T} have noted that there are few, if any, known high-mass DB white dwarfs either in the field or elsewhere. Their Gaia DR2 analysis concluded that DBs of mass greater than 0.8~M$_{\odot}$ are extremely rare and abnormal. A similar result was obtained by \citet{2019MNRAS.486.2169K} on a sample of 1358 DB white dwarfs found in the Sloan Digital Sky Survey. The DB white dwarfs showed a much narrower mass distribution compared to DA white dwarfs, and no DB white dwarfs above 
1.1~M$_{\odot}$ were found. Coupling these results with the presence of a strong magnetic field, the M~47 white dwarf is potentially a Rosetta Stone in understanding the origin of spectral types and magnetic fields in massive white dwarfs. 

\medskip

This work presents results from the European Space Agency (ESA) space mission Gaia. Gaia data are being processed by the Gaia Data Processing and Analysis Consortium (DPAC). Funding for the DPAC is provided by national institutions, in particular the institutions participating in the Gaia MultiLateral Agreement (MLA). The Gaia mission website is https://www.cosmos.esa.int/gaia. The Gaia archive website is https://archives.esac.esa.int/gaia.  

The Pan-STARRS1 Surveys (PS1) and the PS1 public science archive have been made possible through contributions by the Institute for Astronomy, the University of Hawaii, the Pan-STARRS Project Office, the Max-Planck Society and its participating institutes, the Max Planck Institute for Astronomy, Heidelberg and the Max Planck Institute for Extraterrestrial Physics, Garching, The Johns Hopkins University, Durham University, the University of Edinburgh, the Queen's University Belfast, the Harvard-Smithsonian Center for Astrophysics, the Las Cumbres Observatory Global Telescope Network Incorporated, the National Central University of Taiwan, the Space Telescope Science Institute, the National Aeronautics and Space Administration under Grant No. NNX08AR22G issued through the Planetary Science Division of the NASA Science Mission Directorate, the National Science Foundation Grant No. AST-1238877, the University of Maryland, Eotvos Lorand University (ELTE), the Los Alamos National Laboratory, and the Gordon and Betty Moore Foundation.

Based on data products from observations made with ESO Telescopes at the La Silla Paranal Observatory under programme ID 177.D-3023, as part of the VST Photometric H{alpha} Survey of the Southern Galactic Plane and Bulge (VPHAS+, www.vphas.eu).

This research has made use of the VizieR catalogue access tool, CDS,
 Strasbourg, France (DOI : 10.26093/cds/vizier). The original description 
 of the VizieR service was published in \citet{2000A&AS..143...23O}.
 
Based on observations obtained at the Gemini Observatory, which is operated by the Association of Universities for Research in Astronomy, Inc., under a cooperative agreement with the NSF on behalf of the Gemini partnership: the National Science Foundation (United States), National Research Council (Canada), CONICYT (Chile), Ministerio de Ciencia, Tecnología e Innovacion Productiva (Argentina), Ministerio da Ciencia, Tecnologia e Inovacao (Brazil), and Korea Astronomy and Space Science Institute (Republic of Korea). Special thanks to the Canadian Gemini Office for outstanding support in preparation of the observing proposal and its execution.

This work was supported by the Natural Sciences and Engineering Research Council of Canada, the Canadian Foundation for Innovation, the British Columbia Knowledge Development Fund, FRQ-NT (Qu\'ebec) and the National Science Foundation (US).


\bibliography{M47}

\begin{thebibliography}{}
\expandafter\ifx\csname natexlab\endcsname\relax\def\natexlab#1{#1}\fi
\providecommand{\url}[1]{\href{#1}{#1}}

\bibitem[{{Bergh{\"o}fer} {et~al.}(2000){Bergh{\"o}fer}, {Vennes}, \&
  {Dupuis}}]{2000ApJ...538..854B}
{Bergh{\"o}fer}, T.~W., {Vennes}, S., \& {Dupuis}, J. 2000, \apj, 538, 854

\bibitem[{{Briggs} {et~al.}(2018){Briggs}, {Ferrario}, {Tout}, \&
  {Wickramasinghe}}]{2018MNRAS.478..899B}
{Briggs}, G.~P., {Ferrario}, L., {Tout}, C.~A., \& {Wickramasinghe}, D.~T.
  2018, \mnras, 478, 899

\bibitem[{{Burleigh} \& {Jordan}(1998)}]{1998AAS...191.1511B}
{Burleigh}, M.~R., \& {Jordan}, S. 1998, in Bulletin of the American
  Astronomical Society, Vol.~30, American Astronomical Society Meeting
  Abstracts \#191, 1234

\bibitem[{{Camisassa} {et~al.}(2018){Camisassa}, {Althaus}, {C{\'o}rsico}, {De
  Ger{\'o}nimo}, {Miller Bertolami}, {Novarino}, {Rohrmann}, {Wachlin}, \&
  {Garc{\'{\i}}a--Berro}}]{2018arXiv180703894C}
{Camisassa}, M.~E., {Althaus}, L.~G., {C{\'o}rsico}, A.~H., {et~al.} 2018,
  arXiv e-prints, arXiv:1807.03894

\bibitem[{{Chambers} {et~al.}(2016){Chambers}, {Magnier}, {Metcalfe},
  {Flewelling}, {Huber}, {Waters}, {Denneau}, {Draper}, {Farrow}, \&
  {Finkbeiner}}]{2016arXiv161205560C}
{Chambers}, K.~C., {Magnier}, E.~A., {Metcalfe}, N., {et~al.} 2016, arXiv
  e-prints, arXiv:1612.05560

\bibitem[{{Cummings} \& {Kalirai}(2018)}]{2018AJ....156..165C}
{Cummings}, J.~D., \& {Kalirai}, J.~S. 2018, \aj, 156, 165

\bibitem[{{Cummings} {et~al.}(2019){Cummings}, {Kalirai}, {Choi}, {Georgy},
  {Tremblay}, \& {Ramirez-Ruiz}}]{2019ApJ...871L..18C}
{Cummings}, J.~D., {Kalirai}, J.~S., {Choi}, J., {et~al.} 2019, \apjl, 871, L18

\bibitem[{{Cummings} {et~al.}(2016){Cummings}, {Kalirai}, {Tremblay},
  {Ramirez-Ruiz}, \& {Bergeron}}]{2016ApJ...820L..18C}
{Cummings}, J.~D., {Kalirai}, J.~S., {Tremblay}, P.-E., {Ramirez-Ruiz}, E., \&
  {Bergeron}, P. 2016, \apjl, 820, L18

\bibitem[{{Cummings} {et~al.}(2018){Cummings}, {Kalirai}, {Tremblay},
  {Ramirez-Ruiz}, \& {Choi}}]{2018ApJ...866...21C}
{Cummings}, J.~D., {Kalirai}, J.~S., {Tremblay}, P.-E., {Ramirez-Ruiz}, E., \&
  {Choi}, J. 2018, \apj, 866, 21

\bibitem[{{Dobbie} {et~al.}(2006){Dobbie}, {Napiwotzki}, {Lodieu}, {Burleigh},
  {Barstow}, \& {Jameson}}]{2006MNRAS.373L..45D}
{Dobbie}, P.~D., {Napiwotzki}, R., {Lodieu}, N., {et~al.} 2006, \mnras, 373,
  L45

\bibitem[{{Doherty} {et~al.}(2015){Doherty}, {Gil-Pons}, {Siess}, {Lattanzio},
  \& {Lau}}]{2015MNRAS.446.2599D}
{Doherty}, C.~L., {Gil-Pons}, P., {Siess}, L., {Lattanzio}, J.~C., \& {Lau}, H.
  H.~B. 2015, \mnras, 446, 2599

\bibitem[{{Drew} {et~al.}(2014){Drew}, {Gonzalez-Solares}, {Greimel}, {Irwin},
  {K{\"u}pc{\"u} Yoldas}, {Lewis}, {Barentsen}, {Eisl{\"o}ffel}, {Farnhill}, \&
  {Martin}}]{2014MNRAS.440.2036D}
{Drew}, J.~E., {Gonzalez-Solares}, E., {Greimel}, R., {et~al.} 2014, \mnras,
  440, 2036

\bibitem[{{El-Badry} {et~al.}(2018){El-Badry}, {Rix}, \&
  {Weisz}}]{2018ApJ...860L..17E}
{El-Badry}, K., {Rix}, H.-W., \& {Weisz}, D.~R. 2018, \apjl, 860, L17

\bibitem[{{Ferrario}(2018)}]{2018arXiv181006146F}
{Ferrario}, L. 2018, arXiv e-prints, arXiv:1810.06146

\bibitem[{{Ferrario} {et~al.}(2015){Ferrario}, {de Martino}, \&
  {G{\"a}nsicke}}]{2015SSRv..191..111F}
{Ferrario}, L., {de Martino}, D., \& {G{\"a}nsicke}, B.~T. 2015, \ssr, 191, 111

\bibitem[{{Gagn{\'e}} {et~al.}(2018){Gagn{\'e}}, {Fontaine}, {Simon}, \&
  {Faherty}}]{2018ApJ...861L..13G}
{Gagn{\'e}}, J., {Fontaine}, G., {Simon}, A., \& {Faherty}, J.~K. 2018, \apjl,
  861, L13

\bibitem[{{Gaia Collaboration} {et~al.}(2018){Gaia Collaboration}, {Brown},
  {Vallenari}, {Prusti}, {de Bruijne}, {Babusiaux}, {Bailer-Jones}, {Biermann},
  {Evans}, {Eyer}, \& et~al.}]{2018A&A...616A...1G}
{Gaia Collaboration}, {Brown}, A.~G.~A., {Vallenari}, A., {et~al.} 2018, \aap,
  616, A1

\bibitem[{{Gentile Fusillo} {et~al.}(2019){Gentile Fusillo}, {Tremblay},
  {G{\"a}nsicke}, {Manser}, {Cunningham}, {Cukanovaite}, {Hollands}, {Marsh},
  {Raddi}, {Jordan}, {Toonen}, {Geier}, {Barstow}, \&
  {Cummings}}]{2019MNRAS.482.4570G}
{Gentile Fusillo}, N.~P., {Tremblay}, P.-E., {G{\"a}nsicke}, B.~T., {et~al.}
  2019, \mnras, 482, 4570

\bibitem[{{Heyl}(2000)}]{2000MNRAS.317..310H}
{Heyl}, J.~S. 2000, \mnras, 317, 310

\bibitem[{{Kalita} \& {Mukhopadhyay}(2019)}]{2019arXiv190502730K}
{Kalita}, S., \& {Mukhopadhyay}, B. 2019, arXiv e-prints, arXiv:1905.02730

\bibitem[{{Kawka}(2018)}]{2018CoSka..48..228K}
{Kawka}, A. 2018, Contributions of the Astronomical Observatory Skalnate Pleso,
  48, 228

\bibitem[{{Kepler} {et~al.}(2019){Kepler}, {Pelisoli}, {Koester}, {Reindl},
  {Geier}, {Romero}, {Ourique}, {Oliveira}, \& {Amaral}}]{2019MNRAS.486.2169K}
{Kepler}, S.~O., {Pelisoli}, I., {Koester}, D., {et~al.} 2019, \mnras, 486,
  2169

\bibitem[{{Kharchenko} {et~al.}(2005){Kharchenko}, {Piskunov}, {R{\"o}ser},
  {Schilbach}, \& {Scholz}}]{2005A&A...438.1163K}
{Kharchenko}, N.~V., {Piskunov}, A.~E., {R{\"o}ser}, S., {Schilbach}, E., \&
  {Scholz}, R.-D. 2005, \aap, 438, 1163

\bibitem[{{Koester} \& {Reimers}(1981)}]{1981A&A....99L...8K}
{Koester}, D., \& {Reimers}, D. 1981, \aap, 99, L8

\bibitem[{{K{\"u}lebi} {et~al.}(2010){K{\"u}lebi}, {Jordan}, {Nelan},
  {Bastian}, \& {Altmann}}]{2010A&A...524A..36K}
{K{\"u}lebi}, B., {Jordan}, S., {Nelan}, E., {Bastian}, U., \& {Altmann}, M.
  2010, \aap, 524, A36

\bibitem[{Lauffer {et~al.}(2018)Lauffer, Romero, \&
  Kepler}]{10.1093/mnras/sty1925}
Lauffer, G.~R., Romero, A.~D., \& Kepler, S.~O. 2018, Monthly Notices of the
  Royal Astronomical Society, 480, 1547

\bibitem[{{Liebert} {et~al.}(2005{\natexlab{a}}){Liebert}, {Bergeron}, \&
  {Holberg}}]{2005ApJS..156...47L}
{Liebert}, J., {Bergeron}, P., \& {Holberg}, J.~B. 2005{\natexlab{a}}, \apjs,
  156, 47

\bibitem[{{Liebert} {et~al.}(2015){Liebert}, {Ferrario}, {Wickramasinghe}, \&
  {Smith}}]{2015ApJ...804...93L}
{Liebert}, J., {Ferrario}, L., {Wickramasinghe}, D.~T., \& {Smith}, P.~S. 2015,
  \apj, 804, 93

\bibitem[{{Liebert} {et~al.}(2005{\natexlab{b}}){Liebert}, {Wickramasinghe},
  {Schmidt}, {Silvestri}, {Hawley}, {Szkody}, {Ferrario}, {Webbink}, {Oswalt},
  {Smith}, \& {Lemagie}}]{2005AJ....129.2376L}
{Liebert}, J., {Wickramasinghe}, D.~T., {Schmidt}, G.~D., {et~al.}
  2005{\natexlab{b}}, \aj, 129, 2376

\bibitem[{{Marigo} \& {Girardi}(2007)}]{2007A&A...469..239M}
{Marigo}, P., \& {Girardi}, L. 2007, \aap, 469, 239

\bibitem[{{Mermilliod}(1995)}]{1995ASSL..203..127M}
{Mermilliod}, J.-C. 1995, in Astrophysics and Space Science Library, Vol. 203,
  Information \& On-Line Data in Astronomy, ed. D.~{Egret} \& M.~A. {Albrecht},
  127--138

\bibitem[{{Nomoto}(1987)}]{Chandrasekhar}
{Nomoto}, K. 1987, \apj, 322, 206

\bibitem[{{Ochsenbein} {et~al.}(2000){Ochsenbein}, {Bauer}, \&
  {Marcout}}]{2000A&AS..143...23O}
{Ochsenbein}, F., {Bauer}, P., \& {Marcout}, J. 2000, \aaps, 143, 23

\bibitem[{{Raymond} {et~al.}(2003){Raymond}, {Szkody}, {Hawley}, {Anderson},
  {Brinkmann}, {Covey}, {McGehee}, {Schneider}, {West}, \&
  {York}}]{2003AJ....125.2621R}
{Raymond}, S.~N., {Szkody}, P., {Hawley}, S.~L., {et~al.} 2003, \aj, 125, 2621

\bibitem[{{Stello} {et~al.}(2016){Stello}, {Cantiello}, {Fuller}, {Huber},
  {Garc{\'{\i}}a}, {Bedding}, {Bildsten}, \& {Silva
  Aguirre}}]{2016Natur.529..364S}
{Stello}, D., {Cantiello}, M., {Fuller}, J., {et~al.} 2016, \nat, 529, 364

\bibitem[{{Szkody} {et~al.}(2002){Szkody}, {Anderson}, {Ag{\"u}eros},
  {Covarrubias}, {Bentz}, {Hawley}, {Margon}, {Voges}, {Henden}, {Knapp},
  {Vanden Berk}, {Rest}, {Miknaitis}, {Magnier}, {Brinkmann}, {Csabai},
  {Harvanek}, {Hindsley}, {Hennessy}, {Ivezic}, {Kleinman}, {Lamb}, {Long},
  {Newman}, {Neilsen}, {Nichol}, {Nitta}, {Schneider}, {Snedden}, \&
  {York}}]{2002AJ....123..430S}
{Szkody}, P., {Anderson}, S.~F., {Ag{\"u}eros}, M., {et~al.} 2002, \aj, 123,
  430

\bibitem[{{Tremblay} {et~al.}(2019){Tremblay}, {Cukanovaite}, {Gentile
  Fusillo}, {Cunningham}, \& {Hollands}}]{2019MNRAS.482.5222T}
{Tremblay}, P.-E., {Cukanovaite}, E., {Gentile Fusillo}, N.~P., {Cunningham},
  T., \& {Hollands}, M.~A. 2019, \mnras, 482, 5222

\bibitem[{{Tremblay} {et~al.}(2016){Tremblay}, {Cummings}, {Kalirai},
  {G{\"a}nsicke}, {Gentile-Fusillo}, \& {Raddi}}]{2016MNRAS.461.2100T}
{Tremblay}, P.-E., {Cummings}, J., {Kalirai}, J.~S., {et~al.} 2016, \mnras,
  461, 2100

\bibitem[{{Yuan} {et~al.}(2013){Yuan}, {Liu}, \& {Xiang}}]{2013MNRAS.430.2188Y}
{Yuan}, H.~B., {Liu}, X.~W., \& {Xiang}, M.~S. 2013, \mnras, 430, 2188

\bibitem[{{Zhao} {et~al.}(2012){Zhao}, {Oswalt}, {Willson}, {Wang}, \&
  {Zhao}}]{2012ApJ...746..144Z}
{Zhao}, J.~K., {Oswalt}, T.~D., {Willson}, L.~A., {Wang}, Q., \& {Zhao}, G.
  2012, \apj, 746, 144

\end{thebibliography}



\end{document}